\documentclass[submission,copyright]{eptcs}
\providecommand{\event}{QPL 2023}
\providecommand{\publicationstatus}{}

\usepackage{iftex}

\ifpdf
  \usepackage[T1]{fontenc}        
\else
  \usepackage{breakurl}           
\fi

\input{cliffordcs3.sty}

\title{Generators and Relations for $3$-Qubit Clifford+$\CS$ Operators}
\author{Xiaoning Bian and Peter Selinger
  \institute{Dalhousie University}}

\providecommand{\titlerunning}{Generators and Relations for $3$-Qubit Clifford+$\CS$ Operators}
\providecommand{\authorrunning}{Xiaoning Bian and Peter Selinger}

\begin{document}
\maketitle

\begin{abstract}
  We give a presentation by generators and relations of the group of
  $3$-qubit Clifford+$\CS$ operators. The proof roughly consists of two
  parts: (1) applying the Reidemeister-Schreier theorem recursively to
  an earlier result of ours; and (2) the simplification of thousands
  of relations into 17 relations. Both (1) and (2) have been formally
  verified in the proof assistant Agda. The Reidemeister-Schreier
  theorem gives a constructive method for computing a presentation of
  a sub-monoid given a presentation of the super-monoid. To achieve
  (2), we devise an almost-normal form for Clifford+$\CS$
  operators. Along the way, we also identify several interesting
  structures within the Clifford+$\CS$ group. Specifically, we
  identify three different finite subgroups for whose elements we can
  give unique normal forms. We show that the $3$-qubit Clifford+$\CS$
  group, which is of course infinite, is the amalgamated product of
  these three finite subgroups. This result is analogous to the fact
  that the $1$-qubit Clifford+$T$ group is an amalgamated product of
  two finite subgroups.
\end{abstract}

\section{Introduction}

Just like Clifford+$T$ circuits, the class of Clifford+$\CS$ circuits
is universal for quantum computing {\cite{Amy-Glaudell-Ross}}. Here,
$\CS$ denotes the controlled-$S$ gate. Amy, Glaudell, and Ross gave a
characterization of the group of $n$-qubit Clifford+$\CS$ operators,
showing that, up to a trivial condition on the determinant, a matrix
is in this group if and only if it is unitary and its matrix entries
belong to the ring $\Zhi$ {\cite{Amy-Glaudell-Ross}}. As a consequence
of this, or alternatively since the $\CS$ gate is representable as a
Clifford+$T$ circuit with $T$-count $3$, the Clifford+$\CS$ group is a
subgroup of Clifford+$T$; see also
{\cite{beverland2020lower}}. Glaudell, Ross, and Taylor gave a normal
form for 2-qubit Clifford+$\CS$ circuits \cite{Glaudell2021}.  In
{\cite{haah2018codes}}, Haah and Hastings showed how to construct a
fault-tolerant $\CS$-gate via magic state distillation. In
{\cite{garion2020synthesis}}, Garion and Cross described a $\CS$- and
$\CX$-optimal canonical form for the 2-qubit group generated by the
gates $\s{X,T,\CX,\CS}$.

This paper is motivated by the problem of optimizing Clifford+$\CS$
circuits. Like the $T$-gate, the $\CS$-gate is a non-Clifford gate
that is relatively expensive to perform in a fault-tolerant regime,
requiring a magic state to be distilled {\cite{garion2020synthesis}}.
It therefore makes sense to try to minimize the number of $\CS$-gates.
For example, one of the relations we found,
\[
\m{\begin{qcircuit}[scale=0.5]
    \grid{7.50}{0,1}
    \controlledSSd{1.25,0}{1};
    \gate{$K$}{2.75,1};
    \controlledSS{4.00,0}{1};
    \gate{$K$}{5.25,1};
    \controlledSS{6.50,0}{1};
\end{qcircuit}
}
 = 
\m{\begin{qcircuit}[scale=0.5]
    \grid{8.00}{0,1}
    \gate{$S^{\dagger}$}{1.25,1};
    \gate{$K$}{2.75,1};
    \controlledSS{4.00,0}{1};
    \gate{$K$}{5.25,1};
    \gate{$S$}{6.75,1};
\end{qcircuit}
},
\]
can sometimes be used to reduce the $\CS$-count. Although we do not
provide a method for minimizing the $\CS$-count, we solve the
important sub-problem of finding a complete set of relations for
$3$-qubit Clifford+$\CS$ circuits. This guarantees that any $3$-qubit
Clifford+$\CS$ circuit can be transformed into any other equivalent
Clifford+$\CS$ circuit by the repeated application of a finite known
set of relations.

Apart from giving a presentation of the group of $3$-qubit
Clifford+$\CS$ circuits by generators and relations, we also identify
several interesting structures within this group along the
way. Specifically, we identify three different finite subgroups for
whose elements we can give unique normal forms. We show that the
$3$-qubit Clifford+$\CS$ group, which is of course infinite, is the
amalgamated product of these three finite subgroups.

The paper is organized as follows. In Section~\ref{sec:background}, we
provide the necessary background, including the definition of the
Clifford+$\CS$ group. We also recall a presentation of the group of
unitary matrices over the ring $\Zhi$ from our earlier work. In
Section~\ref{sec:presentation}, we state our main result and give an
outline of the proof. In Section~\ref{sec:finite}, we present normal
forms for three finite subgroups of the Clifford+$\CS$ operators, as
well as an almost-normal form for Clifford+$\CS$ operators. In
Section~\ref{sec:amal}, we show that the $3$-qubit Clifford+$\CS$
group is the amalgamated product of these three finite subgroups. This
result is analogous to the fact that the $1$-qubit Clifford+$T$ group
is an amalgamated product of two of its finite subgroups.  In
Section~\ref{sec:doc}, we give a brief overview of the accompanying
Agda code. We conclude the paper with some ideas for future work in
Section~\ref{sec:conclusion}.

\section{Background}
\label{sec:background}

\subsection{Clifford+\texorpdfstring{$\CS$}{CS} operators}

Consider the following unitary operators:
\begin{equation*}\label{eqn-generators}
i,\qquad
K = e^{-i\pi/4} H = \frac{1}{1+i}\zmatrix{cc}{1&1\\1&-1},\qquad
S = \zmatrix{cc}{1&0\\0&i},\qquad
\displaystyle \CZ =
\zmatrix{cccc}{1&0&0&0\\0&1&0&0\\0&0&1&0\\0&0&0&-1}.
\end{equation*}
Here, $i$ is a scalar, namely the usual complex unit. $H$ is the
Hadamard gate, and $K$ is a scaled version of the Hadamard gate. $S$
is sometimes called the \emph{phase gate}, and $\CZ$ is the
controlled-$Z$ gate. When closed under multiplication, identities, and
tensor products, these operators generate the \emph{Clifford group}
(possibly up to scalars, depending which scalars are included in the
Clifford group --- for our purposes, the scalars $\pm 1$ and $\pm i$
are sufficient).

Every operator $U$ obtained in this way is of size $2^n\times 2^n$ for
some natural number $n$, and as usual, we say that $U$ is an operator
on $n$ qubits. We write $\Clifford(n)$ for the group of $n$-qubit
Clifford operators. It is well-known that this group is finite for any
given $n$ (see, e.g., {\cite{selinger2013generators}}), and therefore
not universal for quantum computing. We can obtain a universal gate
set by adding the controlled phase gate
\begin{equation*}
\CS = \zmatrix{cccc}{1&0&0&0\\0&1&0&0\\0&0&1&0\\0&0&0&i}.
\end{equation*}
The resulting operators are called the Clifford+$\CS$ operators, and
we write $\CliffordCS(n)$ for the $n$-qubit Clif\-ford+$\CS$ group.

In this paper, we focus on the case $n=3$. Our goal is to give a
complete presentation of the $3$-qubit Clifford+$\CS$ group in terms
of generators and relations. To ensure that all of our generators are
$8\times 8$-matrices, we introduce the following notations: we write
$\CS_{01}=\CS\otimes I\, , CS_{12}=I\otimes CS\, , K_0 = K \otimes I
\otimes I\, , K_1 = I \otimes K \otimes I\, , K_2 = I \otimes I
\otimes K$, and similarly for $S_0$, $S_1$, and $S_2$, where $I$ is
the $2 \times 2$ identity operator.  We identify the scalar $i$ with
the $8\times 8$-matrix $i\,I$. In the notation for controlled gates,
we use the convention that the target is the last index. For example,
$\CX_{01}$ is a controlled $X$-gate with control qubit $0$ and target
qubit $1$. Note that the controlled $S$- and $Z$-gates are symmetric,
in the sense that $\CS_{jk}=\CS_{kj}$ and $\CZ_{jk}=\CZ_{kj}$, and
therefore the order of indices does not matter for them.

We use the following circuit notations for $K$, $S$, $\CZ$, and $\CS$,
respectively:
\[
  K = \mp{0.3}{\begin{qcircuit}[scale=0.5]
      \grid{2.5}{0}
      \gate{$K$}{1.25,0};
    \end{qcircuit}
  },
  \qquad
  S = \mp{0.3}{\begin{qcircuit}[scale=0.5]
      \grid{2.5}{0}
      \gate{$S$}{1.25,0};
    \end{qcircuit}
  },
  \qquad
  \CZ = 
  \mp{0.3}{\begin{qcircuit}[scale=0.5]
      \grid{2}{0,1}
      \controlled{\dotgate}{1,0}{1};
    \end{qcircuit}
  },
  \qquad
  \CS = 
  \mp{0.3}{\begin{qcircuit}[scale=0.5]
      \grid{2}{0,1}
      \controlledS{1,0}{1};
    \end{qcircuit}
  }.
\]
The $\CZ$ gate is usually denoted
\[
  \mp{0.3}{\begin{qcircuit}[scale=0.5]\grid{2}{0,1}\controlled{\gate{$Z$}}{1,0}{1};\end{qcircuit}},
\]
but since it is symmetric with respect to its two qubits, we prefer
the more symmetric notation shown above. We also use a similar notation for
the $\CS$ gate, except that we label it with an ``$i$\,''. 

We number the qubits from top to bottom, and we write the circuits in
the same order as matrix multiplication, i.e., from right to left. For
example,
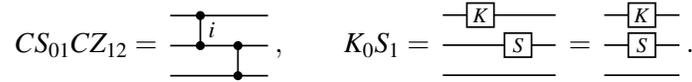
\[
  \CS_{01}\CZ_{12} = \m{\begin{qcircuit}[scale=0.5]
      \grid{3.25}{0,1,2}
      \controlledS{1,1}{2};
      \controlled{\dotgate}{2.25,0}{1};
    \end{qcircuit}
  },
  \qquad
  K_0S_1 = \m{\begin{qcircuit}[scale=0.5]
      \grid{3.75}{0,1,2}
      \invisiblegate{1.25,0};
      \gate{$K$}{1.25,2};
      \gate{$S$}{2.5,1};
    \end{qcircuit}
  } = 
  \m{\begin{qcircuit}[scale=0.5]
      \grid{2.5}{0,1,2}
      \invisiblegate{1.25,0};
      \gate{$K$}{1.25,2};
      \gate{$S$}{1.25,1};
    \end{qcircuit}
  }.
\]
Note that the $X$-gate and the controlled $X$-gate are definable as follows:
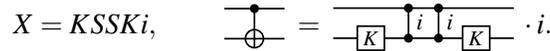
\[
  X = KSSKi, \qquad
  \mp{0.35}{\begin{qcircuit}[scale=0.5]
      \grid{2.00}{0,1}
      \controlled{\notgate}{1.00,0}{1};
    \end{qcircuit}
  }
  =
  \m{\begin{qcircuit}[scale=0.5]
    \grid{6.00}{0,1}
    \gate{$K$}{1.25,0};
    \controlledS{2.50,0}{1};
    \controlledS{3.50,0}{1};
    \gate{$K$}{4.75,0};
  \end{qcircuit}
}\cdot i.
\]
When we use the $X$- and controlled $X$-gates, for example in
Figure~\ref{fig-relations}, they are to be understood as abbreviations
for these definitions.

\subsection{A presentation of \texorpdfstring{$U_n(\Zhi)$}{Uₙ(ℤ[½, i])}}

We briefly recall a result from our earlier work
{\cite{BS2021-gaussian}}.  As usual, $\Z$ is the ring of integers. Let
$\Zhi$ be the smallest subring of the complex numbers containing
$\frac{1}{2}$ and $i$. Let $U_n(\Zhi)$ be the group of unitary
$n\times n$-matrices with entries in $\Zhi$.

In {\cite{BS2021-gaussian}}, we proved that the following is a
presentation of $U_n(\Zhi)$ by generators and relations. The
generators are $i_{[j]}$, $X_{[j,k]}$, and $K_{[j,k]}$, where
$j,k\in\s{0,\ldots,n-1}$ and $j<k$.  The relations are shown in
Figure~\ref{fig:gaussian-relations}. These relations are between words
in the generators, and we write $\eps$ for the empty word
(corresponding to the identity element of the group). The intended
interpretation of the generators is as 1- and 2-level matrices;
specifically, $i_{[j]}$ is like the identity matrix, except with $i$
in the $j$th row and column, and $X_{[j,k]}$ and $K_{[j,k]}$ are like
identity matrices, except with the entries of $X$, respectively $K$,
in the $j$th and $k$th rows and columns, like this:
\[
\def\scale{0.95}
\def\xskip{\hspace{0.3em}}  
i_{[\jay]} \xskip=\xskip
\scalebox{\scale}{$\kbordermatrix{
  & \cdots & \jay & \cdots \\
  \svdots & I & 0 & 0 \\
  \jay & 0 & i & 0  \\
  \svdots & 0 & 0 & I \\
}$},
\quad
X_{[j,k]}\xskip=\xskip
\scalebox{\scale}{$\kbordermatrix{
  &   \dots & j & \dots & k & \dots \\
  \svdots &   I & 0 & 0 & 0 & 0 \\
  j &   0 & 0 & 0 & 1 & 0 \\
  \svdots &   0 & 0 & I & 0 & 0 \\
  k &   0 & 1 & 0 & 0 & 0 \\
  \svdots &   0 & 0 & 0 & 0 & I
}$},
\quad
K_{[j,k]}\xskip=\xskip
\scalebox{\scale}{$\kbordermatrix{
  &   \dots & j & \dots & k & \dots \\
  \svdots &   I & 0 & 0 & 0 & 0 \\
  j &   0 & \frac{1}{1+i} & 0 & \frac{1}{1+i} & 0 \\
  \svdots &   0 & 0 & I & 0 & 0 \\
  k &   0 & \frac{1}{1+i} & 0 & \frac{-1}{1+i} & 0 \\
  \svdots &   0 & 0 & 0 & 0 & I
}$}.
\]

\begin{theorem}[\cite{BS2021-gaussian}]\label{thm:gaussian}
  Let $\Gg$ be the set of one- and two-level matrices $i_{[j]}$,
  $X_{[j,k]}$, and $K_{[j,k]}$, where $j,k\in\s{0,\ldots,n-1}$ and
  $j<k$. Let $\Rr$ be the set of relations shown in
  Figure~\ref{fig:gaussian-relations}. Then $(\Gg,\Rr)$ is a
  presentation of $U_n(\Zhi)$.  In other words, the relations in
  Figure~\ref{fig:gaussian-relations} are sound and complete for
  $U_n(\Zhi)$.
\end{theorem}

\begin{figure}
  \begin{minipage}{0.37\textwidth}
    \begin{align}
      \tli{j}{4} &\,\sim\, \eps \label{eq:1}\\
      \tlx{j}{k}^2 &\,\sim\, \eps \label{eq:2}\\
      \tlk{j}{k}^8 &\,\sim\, \eps \label{eq:3}\\
      \nonumber \\ 
      \tli{j}{}\tli{k}{} & \,\sim\, \tli{k}{}\tli{j}{} \label{eq:4}\\
      \tli{j}{}\tlx{k}{\ell} &\,\sim\,  \tlx{k}{\ell}\tli{j}{}  \label{eq:5}\\
      \tli{j}{}\tlk{k}{\ell} &\,\sim\,  \tlk{k}{\ell}\tli{j}{}  \label{eq:6}\\
      \tlx{j}{k}\tlx{\ell}{m} &\,\sim\, \tlx{\ell}{m}\tlx{j}{k}  \label{eq:7}\\
      \tlx{j}{k}\tlk{\ell}{m} &\,\sim\, \tlk{\ell}{m}\tlx{j}{k}  \label{eq:8}\\
      \tlk{j}{k}\tlk{\ell}{m} &\,\sim\, \tlk{\ell}{m}\tlk{j}{k}  \label{eq:9}
    \end{align}
  \end{minipage}
  \begin{minipage}{0.58\textwidth}
    \begin{align}
      \tli{k}{}\tlx{j}{k} &\,\sim\, \tlx{j}{k}\tli{j}{} \label{eq:10}\\
      \tlx{k}{\ell}\tlx{j}{k} &\,\sim\, \tlx{j}{k}\tlx{j}{\ell} \label{eq:11}\\
      \tlx{\jay}{\ell}\tlx{k}{\ell} &\,\sim\, \tlx{k}{\ell}\tlx{j}{k} \label{eq:11'}\\
      \tlk{k}{\ell}\tlx{j}{k} &\,\sim\, \tlx{j}{k}\tlk{j}{\ell} \label{eq:12}\\
      \tlk{\jay}{\ell}\tlx{k}{\ell} &\,\sim\, \tlx{k}{\ell}\tlk{j}{k} \label{eq:12'}\\
      \nonumber \\ 
    \tlk{j}{k}\tli{k}{2} &\,\sim\, \tlx{j}{k}\tlk{j}{k} \label{eq:13}\\
    \tlk{j}{k}\tli{k}{3}  &\,\sim\, \tli{k}{}\tlk{j}{k}\tli{k}{}\tlk{j}{k} \label{eq:14}\\
    \tlk{j}{k}\tli{j}{}\tli{k}{} &\,\sim\, \tli{j}{}\tli{k}{}\tlk{j}{k} \label{eq:15}\\
    \tlk{j}{k}^2\tli{j}{}\tli{k}{} &\,\sim\, \eps  \label{eq:16}\\
    \tlk{j}{k}\tlk{\ell}{m} \tlk{j}{\ell}\tlk{k}{m}  &\,\sim\,  \tlk{j}{\ell}\tlk{k}{m} \tlk{j}{k}\tlk{\ell}{m}\label{eq:17}
    \end{align}
  \end{minipage}
  \caption{A sound and complete set of relations for $U_n(\Zhi)$. In
    each relation, the indices are assumed to be distinct; moreover,
    whenever a generator $\tlx{a}{b}$ or $\tlk{a}{b}$ is mentioned, we
    assume $a<b$.}\label{fig:gaussian-relations}
  \hfill\rule{0.95\textwidth}{0.1mm}\hfill\hbox{}
\end{figure}

\subsection{The Reidemeister-Schreier theorem}

We will also make use of a result known as the Reidemeister-Schreier
theorem for monoids
{\cite{Reidemeister1927,Schreier1927,BS2022-cliffordt2}}. In a
nutshell, if $G$ is a monoid and $H$ is a submonoid of $G$, the
Reidemeister-Schreier theorem, under suitable assumptions, gives a
method for deriving generators and relations for $H$ from generators
and relations for $G$. Giving a complete account of the
Reidemeister-Schreier theorem is beyond the scope of this paper, but
we refer the interested reader to Section~4.2 of
{\cite{BS2022-cliffordt2}} for a detailed explanation.

\section{A presentation of Clifford+\texorpdfstring{$\CS$}{CS} operators}
\label{sec:presentation}

In this section, we state our main result and give an outline of the
proof. The full proof can be found in the accompanying Agda code
{\cite{Agda-code}}.

\begin{theorem}\label{thm:main}
  The $3$-qubit Clifford+$\CS$ group is presented by $(\Xx,\Ss_X)$, where
  the set of generators is
  \[
    \Xx=\s{i,K_0,K_1,K_2,S_0,S_1,S_2,\CS_{01},\CS_{12}},
  \]
  and the set of
  relations $\Ss_X$ is shown in Figure~\ref{fig-relations}.
\end{theorem}

One interesting feature of the axioms in Figure~\ref{fig-relations} is
that the upside-down version of each relation is also a relation,
except for {\eqref{eq:c15}}. The upside-down version of
{\eqref{eq:c15}} is provable, so we do not require it as an axiom.

\begin{figure}
  \newcounter{mytmpcounter}
  \setcounter{mytmpcounter}{\value{equation}}
  \setcounter{equation}{0}
  \makeatletter
  \renewcommand{\theequation}{C\@arabic\c@equation}
  \makeatother
  \flushleft

  (a) Relations for $n\geq 0$:
  \begin{equation}
    i^4 = \epsilon \label{eq:c1}
  \end{equation}
  (b) Relations for $n\geq 1$:
  \begin{eqnarray}
    K^2 &=& i^3 \label{eq:c2} \\
    S^4 &=& \epsilon \label{eq:c3}\\
    SKSKSK &=& i^3 \label{eq:c4}
  \end{eqnarray}
  (c) Relations for $n\geq 2$:
  \begin{eqnarray}
    \m{\begin{qcircuit}[scale=0.5]
        \grid{5.00}{0,1}
        \controlledS{1.00,0}{1};
        \controlledS{2.00,0}{1};
        \controlledS{3.00,0}{1};
        \controlledS{4.00,0}{1};
      \end{qcircuit}
    }
    &=&
        \m{\begin{qcircuit}[scale=0.5]
            \grid{1.00}{0,1}
          \end{qcircuit}
        } \label{eq:c5}
    \\\nonumber\\[0ex]
\m{\begin{qcircuit}[scale=0.5]
    \grid{3.50}{0,1}
    \gate{$S$}{1.25,1};
    \controlledS{2.50,0}{1};
\end{qcircuit}
     }  
    &=&
 \m{\begin{qcircuit}[scale=0.5]
    \grid{3.50}{0,1}
    \controlledS{1.00,0}{1};
    \gate{$S$}{2.25,1};
\end{qcircuit}
    }\label{eq:c6}
    \\\nonumber\\[0ex]
\m{\begin{qcircuit}[scale=0.5]
    \grid{3.50}{0,1}
    \gate{$S$}{1.25,0};
    \controlledS{2.50,0}{1};
\end{qcircuit}
     }  
    &=&
 \m{\begin{qcircuit}[scale=0.5]
    \grid{3.50}{0,1}
    \controlledS{1.00,0}{1};
    \gate{$S$}{2.25,0};
\end{qcircuit}
    }\label{eq:c7}
    \\\nonumber\\[0ex]
 \m{\begin{qcircuit}[scale=0.5]
    \grid{3.50}{0,1}
    \gate{$X$}{1.25,1};
    \controlledS{2.50,0}{1};
\end{qcircuit}
}
 &=& 
\m{\begin{qcircuit}[scale=0.5]
    \grid{5.50}{0,1}
    \controlledS{1.00,0}{1};
    \controlledS{2.00,0}{1};
    \controlledS{3.00,0}{1};
    \gate{$X$}{4.25,1};
    \gate{$S$}{4.25,0};
\end{qcircuit}
}\label{eq:c8}
    \\\nonumber\\[0ex] 
 \m{\begin{qcircuit}[scale=0.5]
    \grid{3.50}{0,1}
    \gate{$X$}{1.25,0};
    \controlledS{2.50,0}{1};
\end{qcircuit}
}
 &=& 
\m{\begin{qcircuit}[scale=0.5]
    \grid{5.50}{0,1}
    \controlledS{1.00,0}{1};
    \controlledS{2.00,0}{1};
    \controlledS{3.00,0}{1};
    \gate{$X$}{4.25,0};
    \gate{$S$}{4.25,1};
\end{qcircuit}
}\label{eq:c9}
    \\ \nonumber\\[0ex]
\m{\begin{qcircuit}[scale=0.5]
    \grid{7.50}{0,1}
    \gate{$S$}{1.25,1};
    \gate{$K$}{2.75,1};
    \controlledS{4.00,0}{1};
    \gate{$K$}{5.25,1};
    \controlledS{6.50,0}{1};
\end{qcircuit}
     }
    &=&
        \m{\begin{qcircuit}[scale=0.5]
    \grid{7.50}{0,1}
    \controlledS{1.00,0}{1};
    \gate{$K$}{2.25,1};
    \controlledS{3.50,0}{1};
    \gate{$K$}{4.75,1};
    \gate{$S$}{6.25,1};
\end{qcircuit}
}\label{eq:c10}
\\ \nonumber\\[0ex]
\m{\begin{qcircuit}[scale=0.5]
    \grid{7.50}{0,1}
    \gate{$S$}{1.25,0};
    \gate{$K$}{2.75,0};
    \controlledS{4.00,0}{1};
    \gate{$K$}{5.25,0};
    \controlledS{6.50,0}{1};
\end{qcircuit}
     }
     &=&
 \m{\begin{qcircuit}[scale=0.5]
    \grid{7.50}{0,1}
    \controlledS{1.00,0}{1};
    \gate{$K$}{2.25,0};
    \controlledS{3.50,0}{1};
    \gate{$K$}{4.75,0};
    \gate{$S$}{6.25,0};
\end{qcircuit}
}\label{eq:c11}
  \end{eqnarray}
  (d) Relations for $n= 3$:
  \begin{eqnarray}
\m{\begin{qcircuit}[scale=0.5]
    \grid{3.00}{0,1,2}
    \controlledS{1.00,0}{1};
    \controlledS{2.00,1}{2};
\end{qcircuit}
}
 &=& 
\m{\begin{qcircuit}[scale=0.5]
    \grid{3.00}{0,1,2}
    \controlledS{1.00,1}{2};
    \controlledS{2.00,0}{1};
  \end{qcircuit}
               }\label{eq:c12}
    \\\nonumber\\[0ex]
 \m{\begin{qcircuit}[scale=0.5]
    \grid{6.00}{0,1,2}
    \controlled{\notgate}{1.00,2}{1};
    \controlled{\notgate}{2.00,1}{2};
    \controlledS{3.00,0}{1};
    \controlled{\notgate}{4.00,1}{2};
    \controlled{\notgate}{5.00,2}{1};
\end{qcircuit}
}
 &=& 
\m{\begin{qcircuit}[scale=0.5]
    \grid{6.00}{0,1,2}
    \controlled{\notgate}{1.00,0}{1};
    \controlled{\notgate}{2.00,1}{0};
    \controlledS{3.00,1}{2};
    \controlled{\notgate}{4.00,1}{0};
    \controlled{\notgate}{5.00,0}{1};
\end{qcircuit}\label{eq:c13}
} \\ \nonumber\\[0ex]
\m{\begin{qcircuit}[scale=0.5]
    \grid{7.00}{0,1,2}
    \controlledS{1.00,0}{1};
    \controlled{\notgate}{2.00,1}{2};
    \controlledS{3.00,0}{1};
    \controlledS{4.00,0}{1};
    \controlledS{5.00,0}{1};
    \controlled{\notgate}{6.00,1}{2};
\end{qcircuit}
}
 &=& 
\m{\begin{qcircuit}[scale=0.5]
    \grid{7.00}{0,1,2}
    \controlledS{1.00,1}{2};
    \controlled{\notgate}{2.00,1}{0};
    \controlledS{3.00,1}{2};
    \controlledS{4.00,1}{2};
    \controlledS{5.00,1}{2};
    \controlled{\notgate}{6.00,1}{0};
\end{qcircuit}\label{eq:c14}
} \\ \nonumber\\[0ex]
 \m{\begin{qcircuit}[scale=0.5]
    \grid{7.00}{0,1,2}
    \controlled{\notgate}{1.00,2}{1};
    \controlled{\notgate}{2.00,1}{2};
    \controlledS{3.00,0}{1};
    \controlledS{4.00,0}{1};
    \controlled{\notgate}{5.00,1}{2};
    \controlled{\notgate}{6.00,2}{1};
\end{qcircuit}
}
 &=& 
\m{\begin{qcircuit}[scale=0.5]
    \grid{7.00}{0,1,2}
    \controlled{\notgate}{1.00,1}{2};
    \controlledS{2.00,0}{1};
    \controlledS{3.00,0}{1};
    \controlled{\notgate}{4.00,1}{2};
    \controlledS{5.00,0}{1};
    \controlledS{6.00,0}{1};
\end{qcircuit}
     }\label{eq:c15}
    \\ \nonumber\\[0ex]
\m{\begin{qcircuit}[scale=0.5]
    \grid{9.50}{0,1,2}
    \controlledS{1.00,0}{1};
    \gate{$K$}{2.25,1};
    \controlledS{3.50,0}{1};
    \gate{$K$}{4.75,1};
    \controlledS{6.00,1}{2};
    \gate{$K$}{7.25,1};
    \controlledS{8.50,1}{2};
\end{qcircuit}
}
 &=& 
\m{\begin{qcircuit}[scale=0.5]
    \grid{9.50}{0,1,2}
    \controlledS{1.00,1}{2};
    \gate{$K$}{2.25,1};
    \controlledS{3.50,1}{2};
    \gate{$K$}{4.75,1};
    \controlledS{6.00,0}{1};
    \gate{$K$}{7.25,1};
    \controlledS{8.50,0}{1};
\end{qcircuit}
} \label{eq:c16}\\ \nonumber\\[0ex]
 \m{\begin{qcircuit}[scale=0.5]
    \grid{11.50}{0,1,2}
    \controlledS{1.00,0}{1};
    \gate{$K$}{2.25,1};
    \controlledS{3.50,0}{1};
    \controlledS{4.50,0}{1};
    \controlledS{5.50,0}{1};
    \gate{$K$}{6.75,1};
    \controlledS{8.00,1}{2};
    \gate{$K$}{9.25,1};
    \controlledS{10.50,0}{1};
\end{qcircuit}
}
 &=& 
\m{\begin{qcircuit}[scale=0.5]
    \grid{11.50}{0,1,2}
    \controlledS{1.00,1}{2};
    \gate{$K$}{2.25,1};
    \controlledS{3.50,1}{2};
    \controlledS{4.50,1}{2};
    \controlledS{5.50,1}{2};
    \gate{$K$}{6.75,1};
    \controlledS{8.00,0}{1};
    \gate{$K$}{9.25,1};
    \controlledS{10.50,1}{2};
\end{qcircuit}
}\label{eq:c17}\end{eqnarray}

  (e) Monoidal relations: the scalar
  $i$ commutes with everything, and non-overlapping gates commute.

  \caption{Complete relations for $\CliffordCS(3)$. Each relation in
    (b) denotes three relations (one for each qubit), and each
    relation in (c) denotes two relations (one for each pair of
    adjacent qubits).}
  \label{fig-relations}

  \setcounter{equation}{\value{mytmpcounter}}
\end{figure}

\subsection{Proof outline}\label{ssec:outline}

Our proof follows a similar general outline as the corresponding proof
for $2$-qubit Clifford+$T$ operators in {\cite{BS2022-cliffordt2}}.
Let $G=U_8(\Zhi)$ be the group of unitary $8\times 8$-matrices with
entries in $\Zhi$. An exact synthesis algorithm for $G$ was given by
Amy, Glaudell, and Ross {\cite{Amy-Glaudell-Ross}}. Based on this, we
gave a presentation of $G$ by generators and relations in
{\cite{BS2021-gaussian}}.  It is clear that $\CliffordCS(3)$ is a
subgroup of $G$, because all of its generators belong to
$G$. Moreover, by a result of Amy et al.\@ {\cite{Amy-Glaudell-Ross}},
we know that $\CliffordCS(3)$ is precisely the subgroup of $G$
consisting of matrices whose determinant is $\pm 1$. The only other
possible values for the determinant are $\pm i$, and therefore
$\CliffordCS(3)$ is a subgroup of $G$ of index $2$. We can therefore
apply the Reidemeister-Schreier procedure
{\cite{Reidemeister1927,Schreier1927}} to find generators and
relations for $\CliffordCS(3)$, given the known generators and
relations for $G$. Applying this procedure yields a complete set of
relations for $\CliffordCS(3)$.

The application of the Reidemeister-Schreier method produces thousands
of relations, compared to the $17$ cleaned-up relations in
Figure~\ref{fig-relations}. Moreover, these relations are very
large. In our code, which actually uses a sequence of multiple
applications of the Reidemeister-Schreier theorem passing through a
number of intermediate representations, some of the longest relations
involve more than 50,000 generators. Our main contribution is the
simplification of these relations. Due to the sheer magnitude of this
task, we must rely on a computer to expedite the computation.
However, we also require the simplification process to be trustworthy,
as it is very easy in a computer program to accidentally use a
relation that has not yet been proved. To this end, we have formalized
Theorem~\ref{thm:main} and its proof in the proof assistant Agda. This
allows the proof to be verified independently and with a high degree
of confidence in its correctness, despite the magnitude of the proof.

The main idea of the simplification is to use the $17$ relations from
Figure~\ref{fig-relations}, along with some of their easy
consequences, to rewrite the thousands of relations until they are all
eliminated. We define several rewrite systems for this task. Some of
these rewrite systems are confluent and terminating, and others are
just heuristics. All of these rewrite systems are implemented in Agda
and the computations are verified within Agda.

\section{Normal forms and an almost-normal form}
\label{sec:finite}

\subsection{Notations}
\label{ssec:notations}

For convenience, we will use the following notations:
  
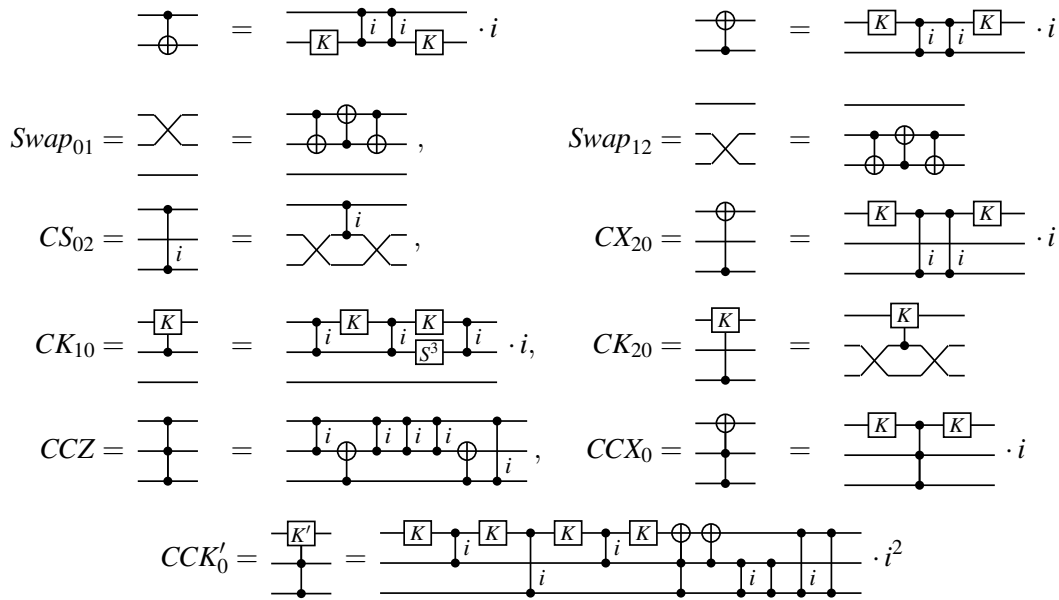
\[
    \begin{array}{rclrcl}
      \m{\begin{qcircuit}[scale=0.5]
          \grid{2.00}{1,2}
          \controlled{\notgate}{1.00,1}{2};
        \end{qcircuit}
      }
      &=& 
          \m{\begin{qcircuit}[scale=0.5]
              \grid{6.00}{1,2}
              \gate{$K$}{1.25,1};
              \controlledS{2.50,1}{2};
              \controlledS{3.50,1}{2};
              \gate{$K$}{4.75,1};
            \end{qcircuit}
          }\cdot i &
                     \m{\begin{qcircuit}[scale=0.5]
                         \grid{2.00}{1,2}
                         \controlled{\notgate}{1.00,2}{1};
                       \end{qcircuit}
                     }
      &=& 
          \m{\begin{qcircuit}[scale=0.5]
              \grid{6.00}{1,2}
              \gate{$K$}{1.25,2};
              \controlledS{2.50,1}{2};
              \controlledS{3.50,1}{2};
              \gate{$K$}{4.75,2};
            \end{qcircuit}
          }\cdot i
      \\\\\nonumber\\[-2ex]          
  \Swap_{01} = \m{\begin{qcircuit}[scale=0.5]
        \grid{2.00}{0,1,2}
        \swapgate{1.00}{1}{2};
      \end{qcircuit}
    }      &=& 
          \m{\begin{qcircuit}[scale=0.5]
              \grid{4.00}{0,1,2}
              \controlled{\notgate}{1.00,1}{2};
              \controlled{\notgate}{2.00,2}{1};
              \controlled{\notgate}{3.00,1}{2};
            \end{qcircuit}
          }\, , &     \Swap_{12} = \m{\begin{qcircuit}[scale=0.5]
        \grid{2.00}{0,1,2}
        \swapgate{1.00}{0}{1};
      \end{qcircuit}
    }
      &=& 
          \m{\begin{qcircuit}[scale=0.5]
              \grid{4.00}{0,1,2}
              \controlled{\notgate}{1.00,0}{1};
              \controlled{\notgate}{2.00,1}{0};
              \controlled{\notgate}{3.00,0}{1};
            \end{qcircuit}
          }    \\\nonumber\\[-1ex]
    \CS_{02} = \m{\begin{qcircuit}[scale=0.5]
        \grid{2.00}{0,1,2}
        \controlledS{1.00,0}{2};
      \end{qcircuit}
    }
      &=& 
          \m{\begin{qcircuit}[scale=0.5]
              \grid{4.00}{0,1,2}
              \swapgate{1.00}{0}{1};
              \controlledS{2.00,1}{2};
              \swapgate{3.00}{0}{1};
            \end{qcircuit}
          }, & \CX_{20} = \m{\begin{qcircuit}[scale=0.5]
              \grid{2.00}{0,1,2}
              \controlled{\notgate}{1.00,2}{0};
            \end{qcircuit}
               }
      &=& 
          \m{\begin{qcircuit}[scale=0.5]
              \grid{6.00}{0,1,2}
              \gate{$K$}{1.25,2};
              \controlledS{2.50,0}{2};
              \controlledS{3.50,0}{2};
              \gate{$K$}{4.75,2};
            \end{qcircuit}
          }\cdot i
    \\\nonumber\\[-1ex]
    \CK_{10} = \m{\begin{qcircuit}[scale=0.5]
        \grid{2.00}{0,1,2}
        \controlled{\gate{$K$}}{1.00,2}{1};
      \end{qcircuit}
    }
      &=& 
          \m{\begin{qcircuit}[scale=0.5]
              \grid{7}{0,1,2}
              \controlledS{1.00,1}{2};
              \gate{$K$}{2.25,2};
              \controlledS{3.50,1}{2};
              \gate{$K$}{4.75,2};
              \controlledS{6.00,1}{2};
              \gate{$S^3$}{4.75,1};
            \end{qcircuit}
          }\cdot i, & 
                      \CK_{20} = \m{\begin{qcircuit}[scale=0.5]
                          \grid{2.00}{0,1,2}
                          \controlled{\gate{$K$}}{1.00,2}{0};
                        \end{qcircuit}
                      }
                    &=& 
                        \m{\begin{qcircuit}[scale=0.5]
                            \grid{4.00}{0,1,2}
                            \swapgate{1.00}{0}{1};
                            \controlled{\gate{$K$}}{2.00,2}{1};
                            \swapgate{3.00}{0}{1};
                          \end{qcircuit}
                        }
    \\\nonumber\\[-1ex]
\CCZ = \m{\begin{qcircuit}[scale=0.5]
    \grid{2.00}{0,1,2}
    \controlled{\dotgate}{1.00,0}{1,2};
\end{qcircuit}
}
 &=& 
\m{\begin{qcircuit}[scale=0.5]
    \grid{8.00}{0,1,2}
    \controlledS{1.00,1}{2};
    \controlled{\notgate}{2.00,1}{0};
    \controlledS{3.00,1}{2};
    \controlledS{4.00,1}{2};
    \controlledS{5.00,1}{2};
    \controlled{\notgate}{6.00,1}{0};
    \controlledS{7.00,0}{2};
\end{qcircuit}
}, & \CCX_0 = \m{\begin{qcircuit}[scale=0.5]
    \grid{2.00}{0,1,2}
    \controlled{\notgate}{1.00,2}{0, 1};
\end{qcircuit}
}
 &=& 
\m{\begin{qcircuit}[scale=0.5]
    \grid{5.00}{0,1,2}
    \gate{$K$}{1.25,2};
    \controlled{\dotgate}{2.50,0}{1,2};
    \gate{$K$}{3.75,2};
\end{qcircuit}
}\cdot i
  \end{array}
     \]

     \[
    \CCK_0 = \m{\begin{qcircuit}[scale=0.5]
        \grid{2.00}{0,1,2}
        \controlled{\gate{$K'$}}{1.00,2}{0, 1};
      \end{qcircuit}
    }
      = 
          \m{\begin{qcircuit}[scale=0.5]
              \grid{16.00}{0,1,2}
              \gate{$K$}{1.25,2};
              \controlledS{2.50,1}{2};
              \gate{$K$}{3.75,2};
              \controlledS{5.00,0}{2};
              \gate{$K$}{6.25,2};
              \controlledS{7.50,1}{2};
              \gate{$K$}{8.75,2};
              \controlled{\notgate}{10.00,2}{0, 1};
              \controlled{\notgate}{11.00,2}{1};
              \controlledS{12.00,0}{1};
              \controlled{\dotgate}{13.00,0}{1};
              \controlledS{14.00,0}{2};
              \controlled{\dotgate}{15.00,0}{2};
            \end{qcircuit}
          }\cdot i^2
\]

The first two notations extend to $3$-qubit circuits, giving us the
definitions of, for example, $CX_{01}$, and $\CX_{21}$. The
definitions for a Toffoli gate with target on the second, respectively
first, qubit are given by
$\CCX_{1} = \Swap_{01}\, \CCX_{0}\, \Swap_{01}$ and
$\CCX_{2} = \Swap_{12}\, \CCX_{1}\, \Swap_{12}$.  The last notation
uses a twice-controlled $K'$ gate. Here $K'=KS^\dagger$ is a variant
of the $K$-gate that has determinant $1$. The reason we are not using
a twice-controlled $K$-gate is that it has determinant $i$ and is
therefore not an element of $\CliffordCS(3)$.

\subsection{Normal forms for finite subgroups of Clifford+\texorpdfstring{$\CS$}{CS} operators}

\begin{figure}
\centering
\begin{tikzpicture}[every node/.style={midway}]
  \matrix[column sep={4em,between origins}, row sep={2em}] at (0,0) {
   & \node(PD) {$PD$}; & \node(KW) {$K_0W$}; &\node(KCD) {$K_0CQD$}; & \\
    \node(P) {$P$}; && \node(CQD) {$CQD$}; && \node(KD) {$K_0QD$}; \\
     \node(W) {$W$}; & \node(CQ) {$CQ$}; && \node(QD) {$QD$}; &\node(K0) {$\gen{K_0}$}; & \\
   & \node(C) {$C$}; & \node(Q) {$Q$}; & \node(D) {$D$}; && \\
  };

  \draw (W) -- (P);
  \draw (K0) -- (KD);
  \draw (K0) -- (KW);
  \draw (W) -- (KW);
  \draw (C) -- (CQ) -- (P) -- (PD);
  \draw (D) -- (QD) -- (KD) -- (KCD);
  \draw (Q) -- (CQ) -- (CQD) -- (KCD);
  \draw (Q) -- (QD) -- (CQD) -- (PD);

\end{tikzpicture}
\caption{The inclusion graph of various finite subgroups of
  $\CliffordCS(3)$}
\label{fig:inclusions}
\end{figure}

We will define normal forms and discuss the structure of the following
finite subgroups of Clifford+$\CS$ operators. The inclusion relations
between these subgroups are visualized in Figure~\ref{fig:inclusions}.

\begin{itemize}
\item $W$, the subgroup of permutation matrices generated by $\Xx_W =
  \s{\Swap_{01},\Swap_{12}}$.
\item $Q$, the subgroup of permutation matrices generated by $\Xx_Q =
  \s{X_0,\CX_{10}, \CX_{20},\CCX_{0}}$.
\item $C$, the subgroup of permutation matrices generated by $\Xx_C =
  \s{X_1,\CX_{12}, \CX_{21}}$.
\item $CQ$, the subgroup generated by $\Xx_C$ and $\Xx_Q$. 
\item $P$, the subgroup of permutation matrices generated by $\Xx_P =
  \s{\CX_{01},\CX_{10},\CX_{12},\CX_{21},\CCX_{0},X_0}$.
\item $D$, the diagonal subgroup generated by $\Xx_D =
  \s{i,S_0,S_1,S_2,\CS_{01},\CS_{12},\CS_{02},\CCZ}$.
\item $PD$, the subgroup generated by $\Xx_P$ and $\Xx_D$. 
\item $QD$, the subgroup generated by $\Xx_Q$ and $\Xx_D$. 
\item $CQD$, the subgroup generated by $\Xx_C$, $\Xx_Q$ and $\Xx_D$. 
\item $K_0D$ the subgroup generated by $\s{K_0} \cup \Xx_D$. Note that
  this group contains $Q$, so it can also be denoted by $K_0QD$.
\item $K_0CD$, the subgroup generated by $\s{K_0} \cup \Xx_C \cup
  \Xx_D$. Since this group contains $Q$, it can also be denoted by
  $K_0CQD$.
\item $K_0W$, the subgroup generated by $K_0$ and $\Xx_W$.
\end{itemize}

Note that $P$ is the group of all permutations of the computational
basis vectors; we call its members ``permutation operators''. $Q$,
$C$, and $CQ$ are subgroups of $P$. Similarly, $D$ is the group of all
diagonal operators in $\CliffordCS(3)$. The remaining subgroups play a
technical role in our proofs.

Given that all claims about finite groups can be proved by just
enumerating the elements, we will not give proofs of the following
claims about finite subgroups of $\CliffordCS(3)$. Instead, we will
illustrate the proofs with examples. Some of the proofs can be found
in the Agda code.

The group $W$ is the group of permutations of $3$ qubits. 

The generators of $Q$ all commute with each other and are
self-inverse. Therefore, each element of $Q$ can be uniquely written
of the form $X_0^a\,\CX_{10}^{b}\,\CX_{20}^{c}\,\CCX_0^{d}$, where
$a,b,c,d\in\s{0,1}$. We say that the subgroup $Q$ has the following
normal form:
\begin{equation}
 \overline{Q} ~~::=~~ X_0^a\,\CX_{10}^{b}\,\CX_{20}^{c}\,\CCX_0^{d}, \text{ where } a,b,c,d \in\s{0,1}. \label{nf-q}
\end{equation}
We use $\overline{Q}$ to range over normal forms for $Q$. More
generally, given any group $G$ for which normal forms are defined, we
use $\overline{G}$ to range over the normal forms of $G$. The group
$Q$ has $2^4=16$ distinct normal forms corresponding to $16$ distinct
elements. 

It is easy to see that $\Swap_{12} \in C$, and therefore also $X_2\in C$.
The group $C$ has the following normal form:
\begin{equation}
  \overline{C} ~~::=~~ c_4c_3c_2 \label{nf-c}
\end{equation}
where
  \[
  \begin{array}{rcl}
    c_2 &\in& \s{\epsilon,\, \CX_{12}},\\
    c_3 &\in& \s{\epsilon,\, \CX_{21},\, \CX_{12}\CX_{21}},\\
    c_4 &\in& \s{X_1^{a}X_2^{b} \mid a,b \in \s{0,1}}.
  \end{array}
  \]
There are $4!=24$ distinct normal forms in $C$.

The group $CQ$ is a semidirect product of $C$ and $Q$ with $Q$ being
normal. A semidirect product structure means that we have commuting
relations of the form $qc=cq'$, or more precisely, for all $q\in Q$
and $c\in C$, there exists a unique $q'\in Q$ such that $qc=cq'$.
Consequently, $CQ$ has the following normal form:
\[
  \overline{CQ} ~~::=~~ \overline{C}\,\,\overline{Q} \label{nf-cq}.
\]

The group $P$ contains $CQ$ as a subgroup with 105 cosets. We get the
following normal form for $P$:
\begin{equation}
  \overline{P} = c\overline{CQ}, \text{ where $c$ ranges over
    the set $V$ of $105$ left coset representatives}. \label{nf-p'}
\end{equation}

One can easily spot a normal form for $D$, since all the generators
commute with each other, $\CCZ$ has order $2$, and all of the other
generators have order $4$. The normal form is:
\begin{equation}
  \overline{D} ~~::=~~
  i^{n_0}S_0^{n_1}S_1^{n_2}S_2^{n_3}\CS_{01}^{n_4}\CS_{12}^{n_5}\CS_{02}^{n_6}\CCZ^{n_7},
  \text{ where } n_0,\ldots,n_6\in\s{0,1,2,3} \text{ and } n_7\in\s{0,1}.
  \label{nf-d}
\end{equation}

The group $PD$ is a semidirect product of $P$ and $D$, with $D$ being
normal. It therefore has the following normal form:
\begin{equation}
  \overline{PD} ~~::=~~ \overline{P}\,\, \overline{D}. \label{nf-pd}
\end{equation}

Since $Q$ is a subgroup of $P$, it follows that $QD$ is also a
semidirect product. It enjoys a similar normal form as 
{\eqref{nf-pd}}, with $P$ replaced by $Q$.

It is easy to see the that group $K_0D$ contains $\Xx_Q$, hence $Q$ is a
subgroup of $K_0D$. We have the following normal form:
\begin{equation}
  \overline{K_0D} ~~::=~~ e_4e_3e_2e_1\overline{D}\,\,\overline{Q}, \label{nf-k0d} 
\end{equation}
where
  \[
  \begin{array}{rcl}
    e_1 &\in& \s{\epsilon,\, \CCK_0,\, \CCK_0\CCK_0},\\
    e_2 &\in& \s{\epsilon,\, \CK_{10},\, S_0\CK_{10}},\\
    e_3 &\in& \s{\epsilon,\, \CK_{20},\, S_0\CK_{20}},\\
    e_4 &\in& \s{\epsilon,\, K_{0},\, S_0K_{0}}.
  \end{array}
\]
Note that $CK_{10},CK_{20}$ and $K_{0}$ commute with each other but
not with $CCK'_{0}$.

Notice that each element of $\Xx_C$ commutes with $K_0$. For any
element of $K_0CD$, for example $w =X_1K_0\CS_{01}K_0\CCZ$, we can
commute $X_1$ all the way to the right using the commuting relations
and the semidirect product structure of $QD$. For example, we get $w =
K_0\CS_{01}\CS_{01}\CS_{01}S_0K_0\CCZ\CS_{02}\CS_{02}X_1$. We will use
the following normal form for $K_0CD$:
\begin{equation}
  \overline{K_0CD} = \overline{(K_0D)}\,\,\,\overline{C} = e_4e_3e_2e_1\overline{D}\,\,\overline{Q}\,\,\overline{C}.\label{nf-k0cd} 
\end{equation}
Note that this also proves that $K_0CD$ is finite, which perhaps
wasn't obvious from its definition.

\subsection{An almost-normal form for \texorpdfstring{$\CliffordCS(3)$}{CS(3)}}

Consider a Clifford+$\CS$ circuit. After replacing the generators
$K_1$ and $K_2$ by $\Swap_{01}\,K_0\,\Swap_{01}$ and
$\Swap_{12}\,\Swap_{01}\,K_0\,\Swap_{01}\,\Swap_{12}$, respectively,
the circuit can be written as an alternating sequence of elements of
$PD$ and $K_0$:
\[
PD\,K_0\,PD\,K_0\ldots PD\,K_0\,PD.
\]
By repeatedly converting subcircuits to normal forms of the form
{\eqref{nf-pd}}, {\eqref{nf-p'}}, and {\eqref{nf-k0cd}}, we can
rewrite this circuit as follows:
\[
\begin{array}{cl}
  & PD\,K_0\,PD\,K_0\ldots PD\,K_0\,PD\\
  \stackrel{\eqref{nf-pd}\eqref{nf-p'}}{\rightarrow}& c \overline{CQ} \overline {D} K_0  c \overline{CQ} \overline {D} K_0 \ldots c \overline{CQ} \overline {D}K_0c \overline{CQ} \overline {D} \\
  \stackrel{\eqref{nf-k0cd}}{\rightarrow} & c e_4e_3e_2e_1\overline{D}\,\,\overline{Q}\,\,\overline{C} c \overline{CQ} \overline {D}K_0\ldots  c \overline{CQ} \overline {D} K_0c \overline{CQ} \overline {D}\\
  \stackrel{\eqref{nf-pd}\eqref{nf-p'}}{\rightarrow} & c e_4e_3e_2e_1 c \overline{CQ} \overline {D}K_0\ldots c \overline{CQ} \overline {D} K_0 c \overline{CQ} \overline {D} \\
  \stackrel{repeat}{\rightarrow} & c e_4e_3e_2e_1\,c e_4e_3e_2e_1\,\ldots \,c e_4e_3e_2e_1  c \overline{CQ} \overline {D}.
\end{array}
\]
We can further rewrite the last expression, for example using
relations in Figure~\ref{fig-vk}. After this step, we might get some
new gates that are not in $V$ or of the form $e_i$. In this case, we
continue with the first arrow step. We repeat the whole process until
there is no further simplification. We call the resulting word an
\emph{almost-normal form}.

It turns out this almost-normal form is ``canonical'' enough. It can
be used to show that a complete set of thousands of relations hold by
rewriting both sides of each relation to almost-normal form. Moreover,
all rewriting rules used to get an almost-normal form are consequences
of the relations in Figure~\ref{fig-relations}. This shows that the
relations in Figure~\ref{fig-relations} are complete.

\begin{figure}
  \def\scale{0.45}
  \def\spacing{-1ex}
  \def\xpacing{\vspace{-1ex}}
  \vspace{-3ex}
  \hspace{0ex}\scalebox{0.9}{\begin{minipage}{1.10\textwidth}
      \begin{eqnarray*}
        \m{\begin{qcircuit}[scale=0.5]
            \grid{6.00}{0,1,2}
            \swapgate{1.00}{1}{2};
            \gate{$K$}{2.25,2};
            \swapgate{3.50}{1}{2};
            \gate{$K$}{4.75,2};
          \end{qcircuit}
        }
        &=& 
        \m{\begin{qcircuit}[scale=0.5]
            \grid{6.00}{0,1,2}
            \gate{$K$}{1.25,2};
            \swapgate{2.50}{1}{2};
            \gate{$K$}{3.75,2};
            \swapgate{5.00}{1}{2};
          \end{qcircuit}
        }
        \\
        \m{\begin{qcircuit}[scale=0.5]
            \grid{4.50}{0,1,2}
            \controlled{\notgate}{1.00,0}{1, 2};
            \gate{$K$}{2.25,2};
            \controlled{\gate{$K$}}{3.50,2}{1};
          \end{qcircuit}
        }
        &=& 
        \m{\begin{qcircuit}[scale=0.5]
            \grid{4.50}{0,1,2}
            \gate{$K$}{1.25,2};
            \controlled{\gate{$K$}}{2.50,2}{1};
            \controlled{\notgate}{3.50,0}{1, 2};
          \end{qcircuit}
        } \label{rel-nc1c} \\ 
        \m{\begin{qcircuit}[scale=0.5]
            \grid{4.00}{0,1,2}
            \controlled{\notgate}{1.00,0}{1, 2};
            \controlled{\gate{$K$}}{2.00,2}{0};
            \controlled{\gate{$K'$}}{3.00,2}{0, 1};
          \end{qcircuit}
        }
        &=& 
        \m{\begin{qcircuit}[scale=0.5]
            \grid{5.00}{0,1,2}
            \controlled{\gate{$K$}}{1.00,2}{0};
            \controlled{\gate{$K'$}}{2.00,2}{0, 1};
            \controlled{\notgate}{3.00,0}{1, 2};
            \controlled{\dotgate}{4.00,1}{2};
          \end{qcircuit}
        } \\ 
        \m{\begin{qcircuit}[scale=0.5]
            \grid{4.00}{0,1,2}
            \controlled{\notgate}{1.00,1}{0, 2};
            \controlled{\gate{$K$}}{2.00,2}{1};
            \controlled{\gate{$K'$}}{3.00,2}{0, 1};
          \end{qcircuit}
        }
        &=& 
        \m{\begin{qcircuit}[scale=0.5]
            \grid{5.00}{0,1,2}
            \controlled{\gate{$K$}}{1.00,2}{1};
            \controlled{\gate{$K'$}}{2.00,2}{0, 1};
            \controlled{\notgate}{3.00,1}{0, 2};
            \controlled{\dotgate}{4.00,0}{2};
          \end{qcircuit}
        } \label{rel-nc4}\\ 
        \m{\begin{qcircuit}[scale=0.5]
            \grid{4.00}{0,1,2}
            \controlled{\notgate}{1.00,0}{1, 2};
            \controlled{\notgate}{2.00,0}{2};
            \controlled{\gate{$K$}}{3.00,2}{1};
          \end{qcircuit}
        }
        &=& 
        \m{\begin{qcircuit}[scale=0.5]
            \grid{4.00}{0,1,2}
            \controlled{\gate{$K$}}{1.00,2}{1};
            \controlled{\notgate}{2.00,0}{1, 2};
            \controlled{\notgate}{3.00,0}{2};
          \end{qcircuit}
        } \\ 
        \m{\begin{qcircuit}[scale=0.5]
            \grid{6.00}{0,1,2}
            \gate{$K$}{1.25,2};
            \controlled{\notgate}{2.50,0}{1, 2};
            \gate{$K$}{3.75,2};
            \controlled{\notgate}{5.00,0}{1, 2};
          \end{qcircuit}
        }
        &=& 
        \m{\begin{qcircuit}[scale=0.5]
            \grid{6.00}{0,1,2}
            \controlled{\notgate}{1.00,0}{1, 2};
            \gate{$K$}{2.25,2};
            \controlled{\notgate}{3.50,0}{1, 2};
            \gate{$K$}{4.75,2};
            \controlled{\notgate}{4.75,0}{1};
          \end{qcircuit}
        }
      \end{eqnarray*}
  \end{minipage}}
  \caption{Some relations used to rewrite words of the form $c
    e_4e_3e_2e_1\,c e_4e_3e_2e_1\,\ldots \,c e_4e_3e_2e_1 $.}
  \label{fig-vk}
\end{figure}

\section{Clifford+\texorpdfstring{$\CS$}{CS} is an amalgamated product of three finite groups}
\label{sec:amal}

Let us first recall the definition of an amalgamated product of two
monoids. For category theorists, this is simply a pushout: Given
monoids $M_1$, $M_2$, and $H$ with morphisms $H\to M_1$ and $H\to
M_2$, the amalgamated product $M_1\amal_H M_2$ is the pushout
\[
\begin{tikzcd}
  H \arrow[r] \arrow[d]
  \arrow[dr, phantom, "\ulcorner", very near end]
  & M_2 \arrow[d, dashed] \\
  M_1 \arrow[r, dashed]
  & M_1\amal_H M_2.
\end{tikzcd}
\]
The amalgamated product of three monoids is defined similarly. Suppose
$M_1$, $M_2$, $M_3$, $H_{12}$, $H_{23}$, $H_{13}$ are monoids with
morphisms $H_{jk}\to H_j$ and $H_{jk}\to H_k$ for all relevant $j$ and
$k$.  Then the amalgamated product $P$ is the colimit of the following
diagram, which generalizes a pushout:
\[
\begin{tikzcd}
  && H_{23} \arrow[dd]\arrow[dr] \\
  & H_{13}\arrow[rr] \arrow[dd]
  && M_3\arrow[dd, dashed]
  \\
  H_{12} \arrow[rr]\arrow[dr]
  && M_2\arrow[dr, dashed]\\
  & M_1\arrow[rr, dashed]
  && P.
\end{tikzcd}
\]
In terms of generators and relations, we have the following situation:
Suppose we have three sets of generators $X$, $Y$, and $Z$, and three
monoid presentations $M_1 = \pres{X\cup Y}{\Gamma_1}$,
$M_2 = \pres{X\cup Z}{\Gamma_2}$, and
$M_3 = \pres{Y\cup Z}{\Gamma_3}$.  We can take $H_{12}=\gen{X}$,
$H_{13}=\gen{Y}$ and $H_{23}=\gen{Z}$, with the obvious maps. Then the
amalgamated product $P$ has the presentation
$\pres{X\cup Y\cup Z}{\Gamma_1\cup\Gamma_2\cup\Gamma_3}$.

In cases where $P$ is an infinite monoid or group, it is remarkable
when $M_1$, $M_2$, and $M_3$ can be chosen to be finite. In that case,
the slogan ``the only relations that hold in $P$ are relations that
hold in a finite submonoid of $P$'' applies.

Using the results of this paper, we can show that $\CliffordCS(3)$ is
an amalgamated product of three finite groups. We choose the sets of
generators as follows:
\[
  \begin{array}{rcl}
    X &=& \s{K_0, i},\\
    Y &=& \s{X_0,X_1,X_2,\CX_{12}, \CX_{21}, \CX_{10}, \CX_{20}, \CCX_0,
          S_0, S_1, S_2, \CS_{01}, \CS_{12}, \CS_{02}, \CCZ, i},\\
    Z &=& \s{\Swap_{01}, \Swap_{12}}.
  \end{array}
\]
One can check that $\gen{X\cup Y}=K_0CQD$, $\gen{X\cup Z} = K_0W$, and
$\gen{Y\cup Z}=PD$. Since $X\cup Y$, $X\cup Z$, and $Y\cup Z$ each
generate a finite subgroup of $\CliffordCS(3)$, all that is left to
show is that each relation of $\CliffordCS(3)$ is a consequence of
relations in one of these three subgroups.

Before we prove this, we must adjust the relations of
Figure~\ref{fig-relations} to fit the new set of generators
$X\cup Y\cup Z$. This requires two adjustments.  First, compared to
the set of generators from Theorem~\ref{thm:main}, a number of new
generators have been added, namely $X_0$, $X_1$, $X_2$, $\CX_{12}$,
$\CX_{21}$, $\CX_{10}$, $\CX_{20}$, $\CCX_0$, $\CS_{02}$, $\CCZ$,
$\Swap_{01}$, and $\Swap_{12}$. For each of these, we must add a
defining relation in terms of the old generators. These relations are
as in Section~\ref{ssec:notations}. Second, the two generators $K_1$
and $K_2$ are no longer used, so where they appear in the relations,
they must now be regarded as abbreviations for the words
$\Swap_{01}\,K_0\,\Swap_{01}$ and
$\Swap_{12}\,\Swap_{01}\,K_0\,\Swap_{01}\,\Swap_{12}$, respectively.
With these adjustments, we still have a sound and complete
presentation of $\CliffordCS(3)$ using the generators $X\cup Y\cup Z$.

Now we must show that each of the relations follows from relations
that hold in $\gen{X\cup Y}$, $\gen{X\cup Z}$, or $\gen{Y\cup Z}$.
Many of the relations, such as \eqref{eq:c1}, \eqref{eq:c3},
\eqref{eq:c5}--\eqref{eq:c9}, and \eqref{eq:c12} are already in one of
the three subgroups, so there is nothing else to show for them. The
remaining relations must be proved individually; here, we give a proof
of \eqref{eq:c16} as a representative example. We have:
\[\begin{array}{rcll}
    \multicolumn{3}{l}{\CS_{12}\,K_1 \,\CS_{12}\,K_1 \,\CS_{01}\,K_1 \,\CS_{01}}\\
    &=& \CS_{12} \,\Swap_{01}\,K_0 \,\Swap_{01} \,\CS_{12} \,\Swap_{01}\,K_0 \,\Swap_{01} \,\CS_{01} \,\Swap_{01}\,K_0 \,\Swap_{01} \,\CS_{01} & \mbox{(1)}\\
    &=& \Swap_{01} \,\CS_{02}\,K_0 \,\CS_{02}\,K_0 \,\CS_{01}\,K_0 \,\CS_{01} \,\Swap_{01} & \mbox{(2)}\\
    &=& \Swap_{01} \,\CS_{01}\,K_0 \,\CS_{01}\,K_0 \,\CS_{02}\,K_0 \,\CS_{02} \,\Swap_{01} & \mbox{(3)}\\
    &=& \CS_{01} \,\Swap_{01}\,K_0 \,\Swap_{01} \,\CS_{01} \,\Swap_{01}\,K_0 \,\Swap_{01} \,\CS_{12} \,\Swap_{01}\,K_0 \,\Swap_{01} \,\CS_{12} & \mbox{(4)}\\
    &=& \CS_{01}\,K_1 \,\CS_{01}\,K_1 \,\CS_{12}\,K_1 \,\CS_{12} & \mbox{(5)}
  \end{array}
\]
Here, steps (1) and (5) use the definition of $K_1$, which is at this
point merely an abbreviation for $\Swap_{01}\,K_0\,\Swap_{01}$. Steps
(2) and (4) uses the relations $\Swap_{01}^2=\epsilon$ and $\Swap_{01}
\,\CS_{12} \,\Swap_{01} = \CS_{02}$ and $\Swap_{01} \,\CS_{01}
\,\Swap_{01} = \CS_{01}$. All three of these relations come from
$\gen{Y\cup Z}$. Step (3) uses the relation $\CS_{02}\,K_0
\,\CS_{02}\,K_0 \,\CS_{01}\,K_0 \,\CS_{01} = \CS_{01}\,K_0
\,\CS_{01}\,K_0 \,\CS_{02}\,K_0 \,\CS_{02}$, which comes from
$\gen{X\cup Y}$. In addition to {\eqref{eq:c16}}, there are a number
of other relations to be proved, but they all follow a similar
pattern.

As mentioned in the introduction, there is an analogous result for the
1-qubit Clifford+$T$ group, which is also an infinite group, and which
is an amalgamated product of two finite subgroups. In this case, the
finite subgroups are the Clifford group and the subgroup of diagonal
and permutation operators, which is generated by $T$ and $X$.

\section{An overview of the accompanying Agda code} \label{sec:doc}

This paper is accompanied by a machine-checkable proof of
Theorem~\ref{thm:main} {\cite{Agda-code}}. It has been formalized in
the proof assistant Agda {\cite{Agda}}. The proof assumes only the
result of {\cite{BS2021-gaussian}}, i.e., the soundness and
completeness of a certain set of relations for $U_n(\Zhi)$. Everything
else is proved from first principles, including, for example, a
complete proof of the version of the Reidemeister-Schreier theorem
that we used.

\paragraph{Verifying the proof.}

Readers who are interested in verifying the proof only need to know
the following: The {\em statement} of Theorem~\ref{thm:main} is
contained in the file {\tt Theorem.agda}, and the final step of the
\emph{proof} of Theorem~\ref{thm:main} is contained in the file {\tt
  Proof.agda}. The reason we separated the statement of the theorem
from its proof is to ensure that the statement assumes as little as
possible: in fact, the file {\tt Theorem.agda} is almost completely
self-contained and only depends on a few definitions concerning
generators, words, indices, and two-level relations. On the other
hand, the proof requires a large number of auxiliary files with
definitions, lemmas, tactics, and more. We checked the proof with Agda
2.6.4, and it took about 120 minutes on our laptop.

\paragraph{Reading the proof.}

For readers who are interested in inspecting our proof, here are some
pointers. The folder {\tt Lib} contains some general-purpose
definitions, such as booleans and natural numbers, and some
definitions and tactics related to monoids and relations. The main
parts of the proof are contained in the folders {\tt Step1} -- {\tt
  Step8}. Each of these steps transforms a set of generators and
relations into an equivalent set of generators and relations,
gradually simplifying the relations. The file {\tt Gate.agda} provides
the definitions for all gates used. The file {\tt CosetNF.agda}
contains definitions related to semidirect products and normal
forms. The final proof witness is contained in the file {\tt
  Proof.agda}.

\section{Conclusion and future work}
\label{sec:conclusion}

The main result of this paper is a presentation of the group of
$3$-qubit Clifford+$\CS$ operators by just $17$ relatively simple
relations. We proved this by a combination of a previous result from
{\cite{BS2021-gaussian}}, the Rei\-de\-mei\-ster-Schrei\-er method, and an
Agda program that simplified several thousand large relations into
the aforementioned $17$ simple ones. Doing this simplification by
brute force would not have been feasible. Instead, we proceeded by
identifying a number of finite subgroups of the Clifford+$\CS$
operators, defining normal forms for these, and then combining them
into carefully chosen rewrite rules. After months of fine-tuning,
these rules eventually reduced the relations to a manageable
size.

Unlike our previous work on generators and relations for $2$-qubit
Clifford+$T$ operators {\cite{BS2022-cliffordt2}}, which used a
\emph{Pauli rotation decomposition} to guide the rewriting, we found
that the analog of the Pauli rotation decomposition, i.e., taking
syllables that are conjugates of the $\CS$ gate under the action of
the Clifford operators, does not work very well. Instead, we were
surprised to find that a more useful decomposition was to take
conjugates of $K_0$ (basically a Hadamard gate) under the action of
diagonal and permutation operators. We may call this the
\emph{Hadamard decomposition} of Clifford+$\CS$. In the process, we
learned many interesting facts about finite subgroups of
Clifford+$\CS$. One of these facts is that the $3$-qubits
Clifford+$\CS$ group is an amalgamated product of three of its finite
subgroups. Concretely, this means that every relation that holds in
this group follows from relations that already hold in some finite
subgroup of Clifford+$\CS$.

This work suggests some interesting directions for future work. Many
of our results about finite subgroups of Clifford+$\CS$ are valid for
$n$ qubits, so one may ask whether our generators and relations can
also be extended to circuits with $4$ or more qubits. Currently, the
limiting factor is the prohibitive computational cost of applying the
Reidemeister-Schreier method to a set of 2-level relations for
$16\times 16$-matrices and then simplifying a massive set of
relations. Perhaps a further study of the finite subgroups of
Clifford+$\CS$ will open up an alternative path to this problem. For
example, one may ask whether the $n$-qubit Clifford+$\CS$ group is an
amalgamated product for all $n$. One may further ask the same question
for the $n$-qubits Clifford+$T$ group or its other subgroups of
interest, such as the Clifford+Toffoli group.

The fact that the Hadamard decomposition turned out to be more useful
than the analog of the Pauli rotation decomposition raises the
question whether our earlier work on Clifford+$T$ could benefit from
the same insight. By applying these lessons, perhaps one can come up
with a simpler complete set of relations. For example, our
Clifford+$T$ axiomatization involved a number of obvious relations and
three ``non-obvious'' ones. We were never able to resolve the
question of whether these non-obvious relations actually follow from
something simpler.

Another intriguing question is whether one can find a unique normal
form for $3$-qubit Clifford+$\CS$ circuits, like the Matsumoto-Amano
normal form for $1$-qubit Clifford+$T$ circuits. We currently only
have an ``almost-normal'' form, but the fact that it efficiently
reduced all of our relations is encouraging.

\bibliographystyle{eptcs}
\bibliography{cliffordcs3}

\begin{thebibliography}{10}
\providecommand{\bibitemdeclare}[2]{}
\providecommand{\surnamestart}{}
\providecommand{\surnameend}{}
\providecommand{\urlprefix}{Available at }
\providecommand{\url}[1]{\texttt{#1}}
\providecommand{\href}[2]{\texttt{#2}}
\providecommand{\urlalt}[2]{\href{#1}{#2}}
\providecommand{\doi}[1]{doi:\urlalt{https://doi.org/#1}{#1}}
\providecommand{\eprint}[1]{arXiv:\urlalt{https://arxiv.org/abs/#1}{#1}}
\providecommand{\bibinfo}[2]{#2}

\bibitemdeclare{misc}{Agda}
\bibitem{Agda}
\emph{\bibinfo{title}{Agda Documentation}}.
\newblock \bibinfo{howpublished}{\url{https://agda.readthedocs.io/}}.
\newblock \bibinfo{note}{Accessed: 2023-03-17}.

\bibitemdeclare{article}{Amy-Glaudell-Ross}
\bibitem{Amy-Glaudell-Ross}
\bibinfo{author}{Matthew \surnamestart Amy\surnameend},
  \bibinfo{author}{Andrew~N. \surnamestart Glaudell\surnameend} \&
  \bibinfo{author}{Neil~J. \surnamestart Ross\surnameend}
  (\bibinfo{year}{2020}): \emph{\bibinfo{title}{Number-theoretic
  characterizations of some restricted {Clifford+$T$} circuits}}.
\newblock {\slshape \bibinfo{journal}{Quantum}} \bibinfo{volume}{4}, p.
  \bibinfo{pages}{252}, \doi{10.22331/q-2020-04-06-252}.
\newblock \bibinfo{note}{Also available from \arxiv{1908.06076}}.

\bibitemdeclare{article}{beverland2020lower}
\bibitem{beverland2020lower}
\bibinfo{author}{Michael \surnamestart Beverland\surnameend},
  \bibinfo{author}{Earl \surnamestart Campbell\surnameend},
  \bibinfo{author}{Mark \surnamestart Howard\surnameend} \&
  \bibinfo{author}{Vadym \surnamestart Kliuchnikov\surnameend}
  (\bibinfo{year}{2020}): \emph{\bibinfo{title}{Lower bounds on the
  non-{Clifford} resources for quantum computations}}.
\newblock {\slshape \bibinfo{journal}{Quantum Science and Technology}}
  \bibinfo{volume}{5}(\bibinfo{number}{3}), p. \bibinfo{pages}{035009},
  \doi{10.1088/2058-9565/ab8963}.

\bibitemdeclare{inproceedings}{BS2021-gaussian}
\bibitem{BS2021-gaussian}
\bibinfo{author}{Xiaoning \surnamestart Bian\surnameend} \&
  \bibinfo{author}{Peter \surnamestart Selinger\surnameend}
  (\bibinfo{year}{2021}): \emph{\bibinfo{title}{Generators and relations for
  {$U_n(\mathbb{Z}[1/2,i])$}}}.
\newblock In: {\slshape \bibinfo{booktitle}{Proceedings of the 18th
  International Conference on Quantum Physics and Logic, QPL 2021, Gdansk,
  Poland}}, {\slshape \bibinfo{series}{Electronic Proceedings in Theoretical
  Computer Science}} \bibinfo{volume}{343}, pp. \bibinfo{pages}{145--164},
  \doi{10.4204/EPTCS.343.8}.

\bibitemdeclare{unpublished}{BS2022-cliffordt2}
\bibitem{BS2022-cliffordt2}
\bibinfo{author}{Xiaoning \surnamestart Bian\surnameend} \&
  \bibinfo{author}{Peter \surnamestart Selinger\surnameend}
  (\bibinfo{year}{2022}): \emph{\bibinfo{title}{Generators and relations for
  2-qubit {Clifford+$T$} operators}}.
\newblock \bibinfo{note}{To appear in {\em QPL 2022}. Available from
  \arxiv{2204.02217}}.

\bibitemdeclare{unpublished}{Agda-code}
\bibitem{Agda-code}
\bibinfo{author}{Xiaoning \surnamestart Bian\surnameend} \&
  \bibinfo{author}{Peter \surnamestart Selinger\surnameend}
  (\bibinfo{year}{2023}): \emph{\bibinfo{title}{Agda code accompanying this
  paper}}.
\newblock \bibinfo{note}{Available from
  \url{https://www.mathstat.dal.ca/~selinger/papers/downloads/cliffordcs3/}}.

\bibitemdeclare{article}{garion2020synthesis}
\bibitem{garion2020synthesis}
\bibinfo{author}{Shelly \surnamestart Garion\surnameend} \&
  \bibinfo{author}{Andrew~W \surnamestart Cross\surnameend}
  (\bibinfo{year}{2020}): \emph{\bibinfo{title}{Synthesis of {CNOT}-dihedral
  circuits with optimal number of two qubit gates}}.
\newblock {\slshape \bibinfo{journal}{Quantum}} \bibinfo{volume}{4}, p.
  \bibinfo{pages}{369}, \doi{10.22331/q-2020-12-07-369}.

\bibitemdeclare{article}{Glaudell2021}
\bibitem{Glaudell2021}
\bibinfo{author}{Andrew~N. \surnamestart Glaudell\surnameend},
  \bibinfo{author}{Neil~J. \surnamestart Ross\surnameend} \&
  \bibinfo{author}{Jacob~M. \surnamestart Taylor\surnameend}
  (\bibinfo{year}{2021}): \emph{\bibinfo{title}{Optimal two-qubit circuits for
  universal fault-tolerant quantum computation}}.
\newblock {\slshape \bibinfo{journal}{npj Quantum Information}}
  \bibinfo{volume}{7}(\bibinfo{number}{1}), p. \bibinfo{pages}{103},
  \doi{10.1038/s41534-021-00424-z}.

\bibitemdeclare{article}{haah2018codes}
\bibitem{haah2018codes}
\bibinfo{author}{Jeongwan \surnamestart Haah\surnameend} \&
  \bibinfo{author}{Matthew~B \surnamestart Hastings\surnameend}
  (\bibinfo{year}{2018}): \emph{\bibinfo{title}{Codes and protocols for
  distilling {$T$}, controlled-{$S$}, and {Toffoli} gates}}.
\newblock {\slshape \bibinfo{journal}{Quantum}} \bibinfo{volume}{2},
  p.~\bibinfo{pages}{71}, \doi{10.22331/q-2018-06-07-71}.

\bibitemdeclare{article}{Reidemeister1927}
\bibitem{Reidemeister1927}
\bibinfo{author}{Kurt \surnamestart Reidemeister\surnameend}
  (\bibinfo{year}{1927}): \emph{\bibinfo{title}{Knoten und {Gruppen}}}.
\newblock {\slshape \bibinfo{journal}{Abhandlungen aus dem Mathematischen
  Seminar der Universit{\"a}t Hamburg}}
  \bibinfo{volume}{5}(\bibinfo{number}{1}), pp. \bibinfo{pages}{7--23},
  \doi{10.1007/BF02952506}.

\bibitemdeclare{article}{Schreier1927}
\bibitem{Schreier1927}
\bibinfo{author}{Otto \surnamestart Schreier\surnameend}
  (\bibinfo{year}{1927}): \emph{\bibinfo{title}{Die {Untergruppen} der freien
  {Gruppen}}}.
\newblock {\slshape \bibinfo{journal}{Abhandlungen aus dem Mathematischen
  Seminar der Universit{\"a}t Hamburg}}
  \bibinfo{volume}{5}(\bibinfo{number}{1}), pp. \bibinfo{pages}{161--183},
  \doi{10.1007/BF02952517}.

\bibitemdeclare{article}{selinger2013generators}
\bibitem{selinger2013generators}
\bibinfo{author}{Peter \surnamestart Selinger\surnameend}
  (\bibinfo{year}{2015}): \emph{\bibinfo{title}{Generators and relations for
  $n$-qubit {Clifford} operators}}.
\newblock {\slshape \bibinfo{journal}{Logical Methods in Computer Science}}
  \bibinfo{volume}{11}(\bibinfo{number}{2:10}), pp. \bibinfo{pages}{1--17},
  \doi{10.2168/LMCS-11(2:10)2015}.
\newblock \bibinfo{note}{Also available from \arxiv{1310.6813}}.

\end{thebibliography}

\end{document}